\documentclass[letterpaper,twocolumn,aps,pra,amssymb,amsfonts,superscriptaddress,showpacs]{revtex4}

\pdfoutput=1
\usepackage{amsmath}
\usepackage{amsfonts}
\usepackage{bm}        
\usepackage{mathrsfs}  
\usepackage{fullpage}
\usepackage[colorlinks,linkcolor=blue,citecolor=blue]{hyperref}
\usepackage{graphicx}
\usepackage{color}

    \newcommand{\eqn}[1]{Eq. (\ref{#1})}       
    \renewcommand{\vec}[1]{\bm{#1}}            

    \def\>{\rangle}                            
    \def\<{\langle}                            %
    
    \newcommand{\R}{R}

\begin{document}

\title{Robustness of composite pulses to time-dependent control noise}

\author{Chingiz Kabytayev}
\affiliation{Schools of Chemistry and Biochemistry; 
Computational Science and Engineering; \\ and Physics,
Georgia Institute of Technology, Atlanta, Georgia 30332-0400, USA}

\author{Todd J. Green}
\affiliation{ARC Centre for Engineered Quantum Systems, School of Physics, 
The University of Sydney, Sydney, New South Wales 2006, Australia}

\author{Kaveh Khodjasteh}
\affiliation{Department of Physics and Astronomy, Dartmouth College, 
Hanover, NH 03755, USA}

\author{Michael J. Biercuk}
\affiliation{ARC Centre for Engineered Quantum Systems, School of Physics, 
The University of Sydney,
Sydney, New South Wales 2006, Australia}

\author{Lorenza Viola}
\affiliation{Department of Physics and Astronomy, Dartmouth College,  
Hanover, New Hampshire 03755, USA}

\author{Kenneth R. Brown}
\email{ken.brown@chemistry.gatech.edu}
\affiliation{Schools of Chemistry and Biochemistry; 
Computational Science and Engineering; \\ and Physics,
Georgia Institute of Technology, Atlanta, Georgia 30332-0400, USA}

\date{\today}

\begin{abstract}
We study the performance of composite pulses in the presence of time-varying control noise on a single qubit.
These protocols, originally devised only to correct for static, systematic errors, are shown to be robust to time-dependent non-Markovian noise in the control field 
up to frequencies as high as $\sim$10\% of the Rabi frequency.  Our study combines a generalized filter-function approach with asymptotic dc-limit 
calculations to give a simple analytic framework for error analysis applied to a number of composite-pulse sequences relevant to nuclear magnetic resonance as well as quantum information experiments.  Results include examination of recently introduced concatenated composite pulses and dynamically corrected gates, demonstrating equivalent first-order suppression of time-dependent fluctuations in amplitude and/or detuning, as appropriate for the sequence in question.  Our analytic results agree well with  numerical simulations for realistic $1/f$ noise spectra with a roll-off to $1/f^2$, providing independent validation of our theoretical insights.
\end{abstract}


\maketitle

\section{Introduction}

High-fidelity control of quantum systems is limited by unwanted interactions with the environment and imperfections in the applied control fields.  Composite pulse (CP) sequences have long been employed 
in nuclear magnetic resonance (NMR) to mitigate the effects of systematic errors in the 
control~\cite{Levitt1981,Wimperis1994,Brown2004,True2012}.  Initially developed to tackle static but otherwise unknown errors in the amplitude or frequency of the driving field, CPs are expressed as the composition of rotations. CPs have been recently extended to handle multiple 
error sources using symmetry \cite{Suter,Odedra} and concatenation \cite{Tomita2010,Bando2013} and to provide efficient high-order error suppresion by optimized design ~\cite{Chuang2013}.  These capabilities have made CPs broadly attractive in laboratory quantum systems, including experimental platforms based on atomic~\cite{AMO} and solid-state ~\cite{Solid,Martinis2003} qubits.  

Despite these advances, an outstanding challenge to the systematic incorporation of CPs into practical quantum information systems remains the limited understanding of CP performance in the presence of 
{\em time-dependent} noise. This is in contrast to optimal control approaches for gate synthesis, where the presence of time-dependent noise is typically assumed in the control design (see, e.g., ~\cite{Montangero2007, Khodjasteh2012, Kosut2013}).
  Previous studies for CPs have examined a restricted set of time-dependent fluctuations in the control including the numeric characterization of decoherence due to random-telegraph noise in the qubit frequency \cite{Mottonen2006}, the effect of stochastic fluctuations in the phase of the control \cite{Zilong2012},  and  the effect of 1/$f^\alpha$ noise on singlet-triplet spin qubits ~\cite{Wang2014}.

Treating the influence of time-dependent noise processes on quantum control operations beyond these limited examples has been facilitated by recent advances in dynamical error suppression based on open-loop Hamiltonian 
engineering~\cite{ViolaDD1,DCG1,Schirmer,ToddNJP}.
These approaches provide a general framework for understanding and mitigating non-Markovian time-dependent noise in a finite-dimensional open quantum system due to either uncontrolled couplings to the environment or a variety of control errors.  In particular, both dynamical decoupling~\cite{ViolaDD1,DD2} and dynamically corrected gates (DCGs)~\cite{DCG1,DCG2, Khodjasteh2012} are able to perturbatively reduce the effects of classical as well as 
quantum noise sources, provided that the correlation time scale of the noise is sufficiently long compared to the control time scale at which the noise is ``coherently averaged out.'' 
These characteristics may be captured quantitatively in filter-transfer functions (FFs) for arbitrary single-qubit control using methods of spectral overlap in the frequency domain~\cite{Kurizki2001,ToddNJP}.
The resulting approach allows for prediction of the leading-order contribution to 
fidelity loss, and has been applied to the study of both dynamically protected 
memory~\cite{MikeExp,Biercuk2011,Viola2013} and nontrivial quantum logic 
operations~\cite{GreenFF, ToddNJP} with results borne out through a variety of 
experiments~\cite{MikeExp,Szwer2010,Soare2014}. 

\begin{table*}[htbp] 
\caption{\label{tab:comp_pulses}CP sequences correcting the target rotation $R(\theta, 0)$ against different error models \cite{True2012,Bando2013}. 
Here, $\phi_1 = \cos^{-1}(-\theta/4\pi)$, $k = \arcsin[\sin(\theta/2)/2]$, $a$, amplitude noise; $d$, detuning noise; $s$, simultaneous amplitude and detuning noise (see text). 
For the DCG sequence \cite{DCG1,ToddNJP}, 
$\Omega_1 = \Omega,\,0\leq t < t_1\equiv \tau/4;$ 
$\Omega_2 = \Omega/2,\,t_1\leq t < t_2\equiv 3\tau/4;$ $\Omega_3 = \Omega,\,t_2\leq t < t_3\equiv \tau$\vspace*{2mm}. }
\begin{tabular}{l c c c c c c c} \hline \hline \begin{tabular}{@{}l@{}}Composite\\pulse\end{tabular}   & \begin{tabular}{@{}c@{}}Error\\model\end{tabular} &   ($\theta_1, \phi_1$) &   ($\theta_2, \phi_2$) &  ($\theta_3, \phi_3$) &  ($\theta_4, \phi_4$) & ($\theta_5, \phi_5$) & ($\theta_6, \phi_6$)   \\
\hline
\hline SK1    & $a$ &($\theta$, 0) &  (2$\pi, -\phi_{1}$)  &  (2$\pi, \phi_{1}$)   & - & - & -\\
\hline BB1    & $a$ &($\theta$, 0) &  ($\pi, \phi_{1}$)  &  (2$\pi, 3 \phi_{1} $) & 
($\pi, \phi_{1}$) & - & - \\
\hline CORPSE    & $d$ &($2\pi + \theta/2 - k, 0$) &  ($2\pi -2k, \pi$) & ($\theta/2 - k, 0$)   
& - & - & - \\
\hline Reduced CinSK  & $s$ & ($2\pi + \theta/2 - k, 0$) &  ($2\pi -2k, \pi$) & ($\theta/2 - k, 0$)   & (2$\pi, -\phi_{1}$) &  (2$\pi, \phi_{1}$) & - \\
\hline Reduced CinBB  & $s$ & ($2\pi + \theta/2 - k, 0$) &  ($2\pi -2k, \pi$) & ($\theta/2 - k, 0$)   & ($\pi, \phi_{1}$)  &  (2$\pi, 3\phi_{1} $)   & ($\pi, \phi_{1}$) \\
\hline \end{tabular}
\end{table*}

In this work, we use a combination of analytic formulations based on FFs and 
numerical simulations to demonstrate that CPs are able to effectively suppress control 
errors caused by time-dependent processes possessing realistic noise power spectra.  
Specifically, we consider a variety of both standard and concatenated CP sequences 
on a single qubit, as well as simple DCG protocols, and compare their performance within a unified 
control framework. Remarkably, robust performance of CP sequences is found {\em up to fluctuations as 
fast as $\sim$10\% of the Rabi frequency}, providing an explicit {\em quantitative} characterization of the sensitivity of 
these approaches to time-dependent control noise.  Calculations show that even under such noise environments, 
which are beyond the static ones originally assumed for CPs, predicted fidelities are at least comparable to those for DCGs in scenarios where protocols of both kind are applicable.  We present a geometric interpretation of CP performance under  time-dependent amplitude noise in order to provide insight into this behavior, further linking the FF formalism with known techniques in CP construction~\cite{True2012}.
 
\section{Theoretical framework} 

\subsection{Control protocols}
\label{subs:protocols}

Both CP and DCG protocols consist of multiple elementary 
control operations, which are sequentially implemented in such a way that 
the desired target operation (quantum gate) is realized 
while simultaneously reducing the net sensitivity to error.  The mathematical
frameworks and error-model assumptions employed in arriving at these
constructions vary considerably, leading to different control modalities. 
While we refer to the relevant literature for a more complete discussion 
\cite{True2012,DCG1,DCG2, Khodjasteh2012},    
we focus here on the task of effecting a target rotation on a
single qubit, which as usual may be parametrized as 
 $$R(\theta,\phi)=\exp[-i  \theta\vec{\rho}(\phi)\cdot\vec{\sigma}/2], \quad 
\vec{\sigma}\equiv (\sigma_{x},\sigma_{y},\sigma_{z}).$$
Ideally, $R(\theta,\phi)$ rotates the qubit Bloch vector through an angle
$\theta$, about an axis defined by the unit vector $\vec{\rho}(\phi) \equiv 
(\cos{\phi}, \sin{\phi},0)$. In practice, any environmental and/or
control errors cause the actual effect of a control protocol to differ from 
the intended one.  We are interested here in error models that may 
be pictured in terms of coupling to {\em classical} degrees of freedom, 
as arise from noisy control actions and/or a fluctuating background 
environment --- in which case the net result is the implementation of a 
different operation on the target system, say, 
$M(\theta,\phi)\neq R(\theta,\phi)$.

The standard error model assumed in CP constructions involves a combination 
of {\em static} (dc) pulse-length and off-resonance control errors, which we may represent as
$$M(\theta,\phi) =
\exp[-i \theta \{ (1 + \epsilon_{a}) \vec{\rho}(\phi)\cdot\vec{\sigma} +
\epsilon_{d} \sigma_{z} \}/2],$$ 
\noindent 
where $\epsilon_{a}$ and $\epsilon_{d}$ quantify 
the amplitude and detuning offsets, respectively. CPs rely on the
application of constant-amplitude control fields segmented into rotations
of different durations about different axes (phase modulation) to counter
these errors which, until recently \cite{Tomita2010,Suter,Odedra,Bando2013}, have 
been addressed separately.  
If $M_{a}(\theta,\phi)$ [respectively, $M_{d}(\theta,\phi)$] denote the
propagator for the special case in which {\em only} $\epsilon_{a}$ 
[respectively, $\epsilon_{d}$] is significant, an $m$th-order CP protocol
$M_{\mu}^{[m]}(\theta,\phi)$ is a sequence of elementary operations
$\{M_{\mu}(\theta,\phi)\}$ for which~\cite{True2012} 
$$M_{\mu}^{[m]}(\theta,\phi) =
  R(\theta,\phi) + \mathcal{O}(\epsilon_{\mu}^{m\mathfrak{}+1}), \quad 
  \mu\in\{a,d\}.$$  
The representative CP sequences we consider are listed in Table \ref{tab:comp_pulses}. 
For instance, SK1 and BB1 are first- and second-order CPs correcting for pure amplitude 
errors~\cite{Brown2004,Wimperis1994}, whereas CORPSE is a first-order compensating 
sequence for pure detuning errors~\cite{Cummins2003}. 
Simultaneous errors can be systematically suppressed 
for arbitrary $(\theta, \phi)$
by applying concatenated CPs~\cite{Bando2013}, such as reduced CinSK (CORPSE in SK1) 
and reduced CinBB (CORPSE in BB1).  

DCG protocols are constructed from general Hamiltonian models for finite-dimensional 
open quantum systems exposed to non-Markovian decoherence 
due to quantum or, as considered here, classical environments.  
This is to be contrasted with CP constructions, which are obtained without making reference 
to an underlying physical model for the intervening error dynamics.
In the simplest case DCGs employ piecewise-constant amplitude and phase modulation 
of the applied control fields across a sequence of carefully designed elementary segments.  
Through this approach, the error sensitivity of the target operation is perturbatively minimized to 
a given order~\cite{DCG1}.  More general analytical DCG constructions are also possible, involving 
``stretching and scaling'' arbitrary control profiles. 
In the present setting, we take advantage of the formal similarity of 
the propagator $M(\theta, \phi)$ under pure off-resonance errors ($\epsilon_a=0$) 
to the one arising from single-axis classical decoherence in the DCG context. 
Specifically, the representative DCG we study is a first-order three-segment sequence, 
obtained from general constructions in the special case  $\theta =\pi$ \cite{DCG1,ToddNJP} 
(see also Table \ref{tab:comp_pulses}).   

\subsection{ Time-dependent error model }
\label{subs:analysis}

In order to both introduce and analyze the effect of \emph{time-dependent} amplitude and 
detuning errors in CP sequences, and compare them to DCGs in a unified setting, it is necessary 
to formulate the control problem at the Hamiltonian (rather than  propagator) level.  
While here we assume piecewise-constant control, more general control modulations may be included as discussed in the Appendix.

Let us consider a piecewise-constant chain of $n$ discrete time-segments, 
each indexed by $l$ and spanning time $t\in[t_{l-1},t_{l}]$ such that, in units of $\hbar \equiv 1$ and 
in a suitable frame, we may write a total Hamiltonian of the form
\begin{eqnarray}
\label{eq:totham}
H(t) &= & \sum_{l=1}^{n} G^{(l)}(t)\frac{\left[\Omega_{l} + \beta_{a}(t)\right]}{2} 
\vec{\rho}^{(l)}_{a}\cdot\vec{\sigma} + \frac{\beta_{d}(t)}{2} \sigma_{z} \nonumber \\ 
     &\equiv & H_0(t) + H_{\text{err}}(t).  
\end{eqnarray}
\noindent
Here, we have introduced a modulation function $G^{(l)}(t)\equiv\Theta[t-t_{l-1}]\Theta[t_{l}-t]$, which has 
unit value for $t\in[t_{l-1},t_{l}]$, and is equal to 0 otherwise, in order to capture the fact that the control is implemented in a piecewise-constant fashion. The ideal control-field amplitude for the $l$-th segment is denoted $\Omega_{l}$, and its axis of rotation, $\vec{\rho}_{a}^{(l)}\equiv\vec{\rho}(\phi_{l})=(\cos{\phi_{l}},\sin{\phi_{l}},0)$. 
The two {\em zero-mean Gaussian} (stationary) stochastic processes 
$\beta_{a}(t)$ and $\beta_{d}(t)$ model amplitude and detuning noise, respectively. We assume that both these processes enter the dynamics additively, and are independent of the ideal amplitude and phase of the 
control, while also being {\em mutually independent}; that is, $\langle\beta_{a}(t)\beta_{d}(t')\rangle=0$.

The total Hamiltonian in Eq. (\ref{eq:totham})  may be separated into ideal 
plus error Hamiltonians by isolating the noise terms proportional to $\beta_\mu$. 
That is, acting alone, $H_{0}(t)$ generates the unitary propagator
$U_{0}(t,0)=\sum_{l=1}^{n}G^{(l)}(t)U_{0}(t,t_{l-1})R'_{l-1}$, which 
describes a sequence of $n$ consecutive elementary control operations $R_{l}\equiv R(\theta_{l},\phi_{l})$, $l=1,\ldots,n$, executed over a total gating time $\tau\equiv t_{n}$. Here, the operator $U_{0}(t,t_{l-1}) \equiv 
\exp[-i \Omega_{l}(t - t_{l-1}) \vec{\rho}_{a}^{(l)}\cdot\vec{\sigma}/2]$ is the time-dependent propagator for the $l$-th elementary pulse, such that $\theta_{l} = \Omega_{l} \,(t_{l} - t_{l-1})$ and $U_{0}(t_{l},t_{l-1})=R_{l}$. At the end of the sequence, $U_{0}(\tau,0)=R(\theta,\phi)=R'_{n}$ (the desired target operation), where $R'_{l}\equiv\R_{l}R_{l-1}\ldots R_{0}$ and $R_{0}\equiv I$.

Following~\cite{DCG1}, the total evolution operator $U(t,0)$, generated by the 
controlled Hamiltonian in Eq. (\ref{eq:totham}), may then be written as 
$U(\tau,0) \equiv U_{0}(\tau,0)\exp[-i\Phi(\tau)]$, where the ``error action operator'' 
encapsulates the effect of $H_{\text{err}}(t)$ and, 
to the lowest order in a perturbative Magnus-series expansion, we may write 
\begin{equation}
\Phi(\tau)\approx \Phi_1(\tau)=\int_{0}^{\tau}dt\,U_{0}^{\dag}(t,0)H_{\text{err}}(t)U_{0}(t,0).
\label{eq:first}
\end{equation}
Calculating this quantity requires consideration of all (ideal) time-ordered control operations 
enacted during the sequence; accordingly, let us define ``control vectors'' as \cite{ToddNJP}: 
\begin{eqnarray*} 
\vec{\rho}_{a}(t)&\equiv&\frac{1}{2}\sum_{l=1}^{n}G^{(l)}(t)\vec{\rho}_{a}^{(l)}\mathbf{\Lambda}^{(l-1)}, \\ 
\vec{\rho}_{d}(t)& \equiv &\frac{1}{2}\sum_{l=1}^{n}G^{(l)}(t)\vec{\rho}^{(l)}_{d}(t-t_{l-1})\mathbf{\Lambda}^{(l-1)}, 
\end{eqnarray*}
where the matrices (vectors) 
$\mathbf{\Lambda}^{(l-1)}$ [$\vec{\rho}^{(l)}_{d}(t-t_{l-1})$] have components
\begin{eqnarray*}
\Lambda_{ij}^{(l-1)} & =& \text{Tr}[R'^{\dag}_{l-1}\sigma_{i}R'_{l-1}\sigma_{j}]/2 ,\\
{\rho}^{(l)}_{d,j}(t-t_{l-1})&=&\text{Tr}[U_{0}^{\dag}(t,t_{l-1})\sigma_{z}U_{0}(t,t_{l-1})
\sigma_{j}]/2 ,
\end{eqnarray*}
for $i,j\in\{x,y,z\}$. Thus, $\Phi_{1}(\tau)=\vec{a}(\tau)\cdot\vec{\sigma}$, 
where the ``error vector,'' 
$$\vec{a}(\tau)\equiv\int_{0}^{\tau}dt \, [\beta_{a}(t)\vec{\rho}_{a}(t)+\beta_{d}(t)\vec{\rho}_{d}(t)],$$ 
captures the difference between the actual and the target control operations, for each realization of the noise.

In order to gain useful information about the overall performance of a CP protocol, 
we must characterize performance across an ensemble of noise realizations.   As 
a figure of merit, we consider the ensemble-averaged (denoted $\langle\cdot\rangle$) 
propagator fidelity, which, in our qubit setting, reads
$$\mathcal{F}=\frac{1}{4} \left\langle|\text{Tr}[U^{\dag}_{0}(\tau,0)U(\tau,0)]|^2 \right\rangle . $$
In the (weak-noise and/or short-time) limit where the first-order description of Eq. (\ref{eq:first}) 
is accurate, we may further write~\cite{ToddNJP,Soare2014}:
$$\mathcal{F}\approx1-\langle a(\tau)^2\rangle,\quad a(\tau)\equiv [\vec{a}(\tau)\cdot\vec{a}(\tau)]^{1/2}.$$ 
\noindent 
This quantity is most conveniently calculated in the Fourier domain; introducing the noise power 
spectral densities,
\begin{equation*}
S_{\mu}(\omega)\equiv\int_{-\infty}^{\infty}dt \, e^{-i\omega t}\langle\beta_{\mu}(t_0)\beta_{\mu}(t_0+t)\rangle
\end{equation*} 
for $\mu\in\{a,d\}$, and exploiting the stationarity and independence properties of the noise sources, 
we finally obtain the following expression for the (first-order) fidelity loss:
\begin{equation}
\label{eq:fidloss}
1-\mathcal{F} \approx \frac{1}{2\pi}\int_{-\infty}^{\infty}\frac{d\omega }{\omega^2}
\sum_{\mu=a,d}S_{\mu}(\omega)F_{\mu}(\omega).
\end{equation}
Here, $F_{\mu}(\omega) \equiv \vec{\rho}^{*}_{\mu}(\omega)\cdot\vec{\rho}_{\mu}(\omega)$ 
is the {\em generalized FF} for amplitude ($\mu=a$) and detuning ($\mu=d$), respectively, 
defined in terms of the frequency-domain control vectors,
$\vec{\rho}_{\mu}(\omega) = -i\omega\int_{0}^{\tau}dt\vec{\rho}_{\mu}(t)e^{i\omega t}$.

The FFs characterize the spectral properties of the applied control and thus provide a simple quantitative 
means to compare the control protocols of interest (see Table~\ref{tab:comp_pulses}) in the presence of 
time-dependent Gaussian noise~\cite{Biercuk2011,ToddNJP}. In general, one may interpret these functions 
by considering the transfer function of a high-pass filter, including passband, stopband, and roll-off.  The filter 
roll-off, captured by the slope of the FF near zero frequency, serves as a lower bound on the order of error suppression in the presence of time-dependent noise \cite{Paz-Silva2014}. This approach has been validated for nontrivial control --- including CP constructions --- in recent experiments \cite{Soare2014}.  We next proceed to calculate and present 
independently the FFs for both amplitude and detuning quadratures.

\section{Results} 
\begin{figure*}[htbp]
\begin{center}
\includegraphics[width=\textwidth]{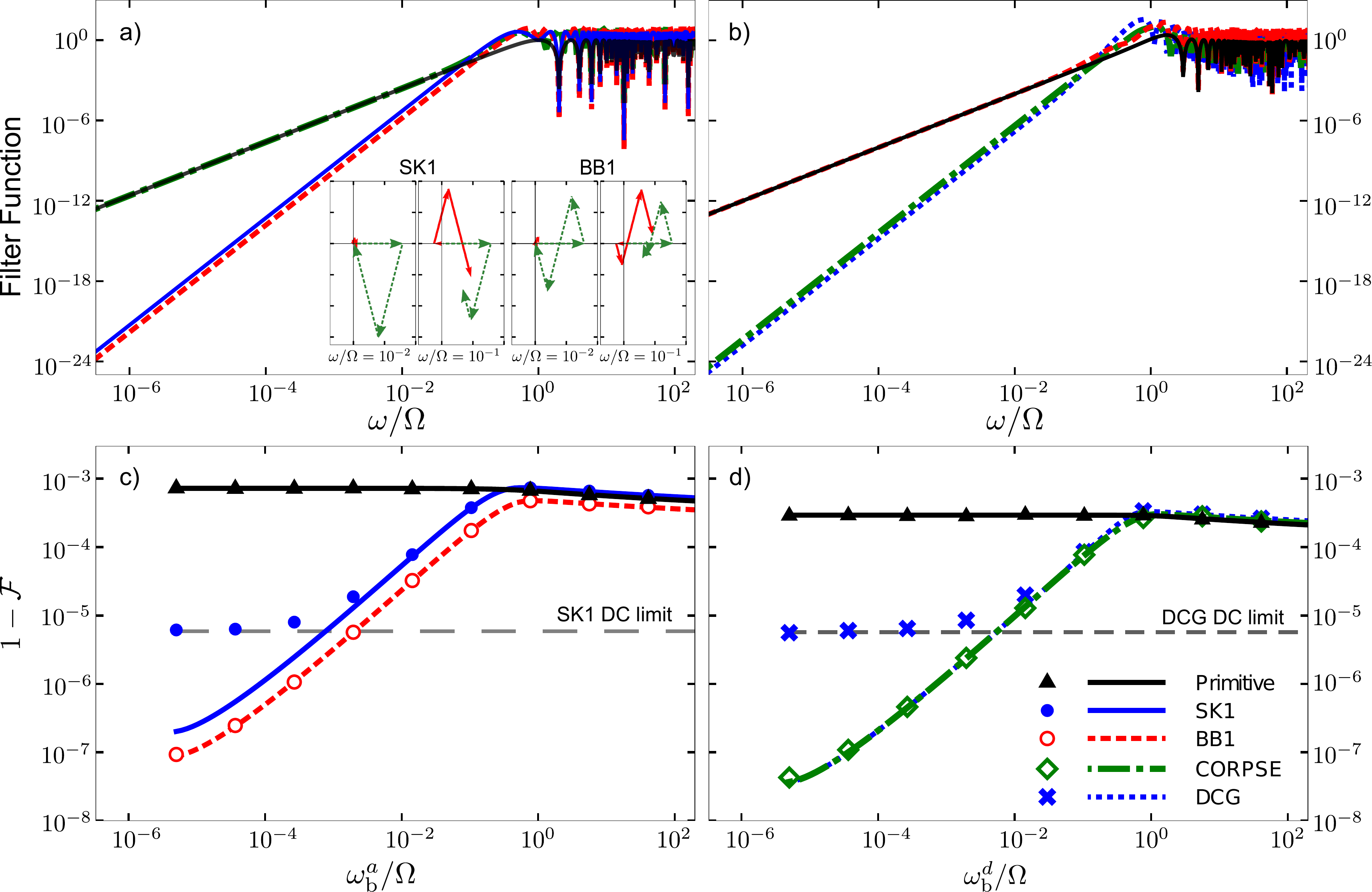}
\end{center}
\vspace*{-4mm}
\caption{(Color online) (a, b) FFs as a function of dimensionless frequency for amplitude (a), and detuning noise (b). A target rotation angle $\theta=\pi$ is used for all sequences. 
Inset: Geometric picture of first- vs. second-order amplitude-error CPs. Axes represent Cartesian $x$ and $y$, indicating the rotation axes of the different segments (see text).  The initial rotation is effected about the $x$-axis, indicated by the horizontal line, with corrections about different axes conducted subsequently.  Returning to the origin indicates suppression of error, with two time-domain elements of the FF being indicated by color (dashed green and solid red). 
(c, d) Performance of CP sequences in the presence of a constant power amplitude (c) and detuning noise (d) with a $1/f$ Gaussian noise spectrum and $1/f^{2}$ roll-off, Eq. (\ref{eq:spec_with_rolloff}).
Spectrum parameters: $A_a = A_d \equiv A = 2.07 \cdot 10^{9}/[\ln(\omega_{\text{b}}/\omega_{\text{min}}) + 1 - (\omega_{\text{b}}/\omega_{\text{max}})]$ (rad/s)$^3$/Hz, where $\omega_{\text{b}}$ is the knee of the roll-off; $\omega_{\text{min}} = 2\pi$ $\text{rad}/\text{s}$, $\omega_{\text{max}} = 4.5 \cdot 10^{9} $ $\text{rad}/\text{s}$. 
Control amplitude: $\Omega = 1.5 \cdot 10^6 \text{rad}/\text{s}$. Numerical simulation involves discretization of the continuous noise functions $\beta_{\mu}(t)$, calculating a single instance of $U(\tau,0)$ and a single value for fidelity, and averaging over $N$ noise realizations. 
We employ the Karhunen-Loeve filter ~\cite{ghanem2012} to simulate discrete noise 
in the Gaussian limit~\cite{Kasdin1995}. 
Analytical lines representing the fidelity loss calculated by the FF approach [Eq. (\ref{eq:fidloss}), in color] 
and by the dc-limit approach [Eq. (\ref{eq:slow}), gray] are plotted. The dc limit for BB1 and CORPSE are below the bounds of the plot at $1-\mathcal{F} = 3.9 \times 10^{-9}$ and $1-\mathcal{F} = 3.0 \times 10^{-9}$, respectively}
\label{fig:pulseFF}
\end{figure*}

\subsection{Analytical results and geometric picture} 

We begin by analyzing the effect of a single noise source, as described by the appropriate generalized 
FF introduced in the previous section.  Results are summarized in Figs. \ref{fig:pulseFF}(a) and \ref{fig:pulseFF}(b) , where we also show, for comparison, FFs for an uncorrected  (elementary or ``primitive'') $\pi$ rotation.  As the latter is expected to have no error-suppressing properties, 
a comparison of the FFs for CP protocols against the primitive rotation reveals their relative performance advantages;  a steeper slope indicates improved (higher-order) error suppression.  All compensating sequences show the expected first-order suppression of errors 
against which they are designed to be effective, in the low-frequency limit.  At the same time, 
they show no improvement over the primitive for the uncompensated error quadrature.  
Remarkably, our analysis reveals that the crossover frequency at which the FF for CP protocols becomes larger 
than that for the primitive is as high as $\sim \% 10$ of the driving frequency $\Omega$. 
Accordingly, in circumstances where the noise power spectral density is 
dominated by frequencies below this value, CP sequences are still expected to provide robust 
error-suppressing performance.  

For amplitude noise, it is possible to make connections between the form of the amplitude FF and geometric 
models commonly used to describe CPs \cite{True2012,Odedra,Jones2013,Chuang2013}. One may represent a compensating sequence as an initial target rotation, followed by correction rotations, captured through a set of vectors in a multi-dimensional space.   Direct calculation shows that a sequence correcting dc errors to the 
first order satisfies the condition: 
\begin{equation}
\sum_l \Omega(t_l-t_{l-1}) \tilde{\vec{\rho}}_{a}^{(l)}= 0, \quad 
\tilde{\vec{\rho}}_{a}^{(l)}\equiv\vec{\rho}_{a}^{(l)}\boldsymbol{\Lambda}^{(l-1)}.
\label{eq:fo}
\end{equation}
\noindent 
If one treats each term in the above sum as a vector of length $\Omega(t_l-t_{l-1})$ pointing in the direction 
$\tilde{\vec{\rho}}_{a}^{(l)}$, then placing the vectors end to end forms a {\em closed} figure, demonstrating the effective DC error suppression.  In this picture, SK1 yields a triangle, whereas BB1 corresponds to two triangles with opposite-signed area, indicating second-order correction, as expected~\cite{True2012}.

Returning to the FF construction, we find that the amplitude-noise FF may be written as
\begin{equation*}
F_{a}(\omega)=\frac{1}{4}\Big\{\Big|\sum_{l}A_{l}(\omega)\tilde{\vec{\rho}}_{a}^{(l)}\Big|^2
+\Big|\sum_{l}B_{l}(\omega)\tilde{\vec{\rho}}_{a}^{(l)}\Big|^2 \Big\},
\end{equation*}
\begin{eqnarray*}
&& A_{l}(\omega)  \equiv \cos(\omega t_{l})-\cos(\omega t_{l-1}) ,\\
&& B_{l}(\omega)  \equiv  \sin(\omega t_{l})-\sin(\omega t_{l-1}). 
\end{eqnarray*} 
The above expression for $F_a(\omega)$ may be interpreted in terms of 
the magnitudes of two three-dimensional real vectors $\vec{A}\equiv 
\sum_{l}A_{l}(\omega)\tilde{\vec{\rho}}_{a}^{(l)}$ and 
$\vec{B}\equiv \sum_{l}B_{l}(\omega)\tilde{\vec{\rho}}_{a}^{(l)}$.   
When $\omega$ is 
small compared to the relevant time scales, Taylor-expansion of $B_l$ shows 
that to second order in $\omega$, we have 
$\sum_{l}B_{l}(\omega)\tilde{\vec{\rho}}_{a}^{(l)} \approx \frac{\omega}{\Omega} \sum_l \Omega(t_l-t_{l-1}) \tilde{\vec{\rho}}_{a}^{(l)}= 0,$ which corresponds to (a scaled version of) the closed-loop condition 
required for error suppression at dc, Eq. (\ref{eq:fo}). To second order, $A_l \approx \frac{\omega^2}{2}(t_l^2-t_{l-1}^2)$,  which thus dominates the error. This implies that all CPs for amplitude noise should have FFs that scale at least as $\omega^4$ in the limit of small $\omega$.  These observations tie to previous knowledge about general FFs and associated error-suppressing properties ~\cite{Biercuk2011,ToddNJP}.

The inset in Fig. \ref{fig:pulseFF}(a) shows the vectors $\vec{A}$ and $\vec{B}$ 
divided by $\omega$ [dashed (green) arrows corresponding to $B_l/\omega$ and solid (red) arrows 
corresponding to $A_l/\omega$] and placed end to end for SK1 and BB1, for two values of the dimensionless frequency $\omega/\Omega$.  At sufficiently small $\omega$ 
the dashed (green) arrows trace an approximate closed path, whereas 
for higher frequencies,  $\omega \gtrsim 0.1 \,\Omega$, 
higher-order terms become important.  In this case, the resulting figure is no longer closed and 
the sequence will not be error suppressing, in agreement with the FF analysis presented above.  
Thus, this geometric picture reflects common observations for dc error analyses, 
but now lifted to a time-dependent error model, analyzed in the frequency domain. 

We can also use the small-$\omega$ limit of $\vec{A}$ and $\vec{B}$ to estimate the crossover frequency at which the CP FF, $F_a^{CP}(\omega)$, will exceed the primitive pulse FF, $F_a^{P}(\omega)$. The primitive pulse FF is determined by the leading term in $\vec{B}$, $F_a^{p}(\omega) \approx \frac{1}{4} (\omega \tau_P)^2$, where $\tau_P$ is the pulse duration. The low frequency CP FF is determined by the leading term in $\vec{A}$, which can be bounded from above by making the assumption that all $\tilde{\vec{\rho}}_{a}^{(l)}$ are the same. This results in $F_a^{CP} =  \frac{1}{16} (\omega \tau_{CP})^4 $, where $\tau_{CP}$ is the length of the CP.  For SK1 and BB1 with $\theta~=~\pi$, $\tau_{CP} = (4\pi + \theta)/\Omega$ and this bound predicts that the CP will reduce the error, $F_a^{CP}(\omega)<F_a^{P}(\omega)$,  when $\omega < 0.025 \Omega$. This is an approximate lower bound; the actual crossover frequencies are $\omega = 0.069 \Omega$ for SK1 and $\omega = 0.127 \Omega$ for BB1.

While these approaches capture the effects of dynamic control noise well, the first-order FF formalism {\em underestimates} error in the region $\omega/\Omega \ll1$, corresponding to noise processes fluctuating slowly on the scale of operation time.  
This may be understood by treating very slow noise as a {\em constant} error term equal 
to the strength of $H_\text{err}$ at the start of the sequence, $\beta_\mu (0)$. For small, constant noise an $m$th-order CP (or DCG) sequence is
well approximated by $U^{[m]}(\tau,0)\approx U_0(\tau,0)\exp[-i\Phi_{m+1}(\tau)]$, where $\Phi_{m+1}(\tau)$ is 
the $(m+1)$th order term in the perturbative Magnus expansion \cite{DCG1,True2012}. 
For a qubit like we consider, $\Phi_{m+1}(\tau)$ is traceless with eigenvalues $\pm \lambda_{m+1}$ and the magnitude of $\lambda_{m+1}$ is proportional to $\beta_\mu(0)^{m+1}$. The fidelity of the sequence is 
then $\mathcal{F}\approx\<\cos(\lambda_{m+1})^2\>$. In this limit, the leading order error term can 
thus be written as
\begin{equation}
\label{eq:slow}
1-\mathcal{F} \approx \< \lambda_{m+1}^2 \> = c_{m+1} \< \beta_\mu(0)^{2(m+1)} \> \\ ,
\end{equation}
where the proportionality constant $c_{m+1}$, like $F_\mu(\omega)$, depends on the 
sequence and the noise axis, but not the noise strength. 

As an example, consider SK1 with constant noise $\beta_a(0)$. The leading-order Magnus term is  
\begin{eqnarray*}
\Phi_2(\tau) &=& i\frac{\beta_a(0)^2}{2} \int^\tau_0 dt \int^t_0 dt^\prime [ \rho_a(t),\rho_a(t^\prime) ]\\
&=&  \beta_a(0)^2\pi^2\sin(2\phi_1) \sigma_z
\end{eqnarray*}
where $\phi_1=\cos^{-1}(-1/4)$ (see Table \ref{tab:comp_pulses}). The eigenvalues of $\Phi_2$ are  $\pm \lambda_2=\pm \beta_a(0)^2\pi^2\sin(2\phi_1)$, and as a result, $1-\mathcal{F}\approx (\pi^2\sin(2\phi_1))^2\<\beta_a(0)^4\>$. The term  $c_2=(\pi^2\sin(2\phi_1))^2$ depends only on the pulse sequence and $ \< \beta_a(0)^4\>$ is averaged over the ensemble of initial noise strengths.

The error of the first-order fidelity approximation in the FF formalism [Eq. (\ref{eq:fidloss})] 
depends only on the first-order Magnus term [Eq. (\ref{eq:first})], so the slow-noise (dc) limit 
contains fidelity loss contributions from higher-order FF terms that are ignored in the first-order 
approximation (see also~\cite{ToddNJP} for additional details). For a zero-mean Gaussian noise 
described by a spectral density $S(\omega)$,  by definition $\< \beta_\mu(0)^2 \> =\int_{-\infty}^{\infty}d\omega S_\mu(\omega)$. All odd orders of the expectation value are 0 and all even orders are proportional to powers of the second order expectation value,
$$ \langle \beta_\mu(0)^{2(m+1)}
\rangle  = (2m+1)!! \Big(\int_{-\infty}^{\infty}d\omega S_\mu(\omega) \Big)^{m+1}.$$
\noindent 
We may therefore estimate the analytical fidelity loss over the entire frequency range by
combining the contributions from Eq. (\ref{eq:fidloss}) and Eq. (\ref{eq:slow}).

\subsection{Comparison with numerical results}

Quantifying the fidelity loss [Eq. (\ref{eq:fidloss})] for control protocols implemented in a real 
(classical) noise environment requires one to choose a specific noise spectrum. As a practical example, 
we consider $1/f$ Gaussian noise with a roll-off to $1/f^2$ noise at high frequency with spectrum 
\begin{equation} S_{\mu}(\omega) =  \left\{ 
\begin{array}{ccc}
A_{\mu}/\omega, & \quad \omega^{\mu}_{\text{min}} < \omega < \omega^{\mu}_{\text{b}}, \\ 
\omega^{\mu}_{\text{b}} \cdot A_{\mu}/\omega^2, & \quad \omega^{\mu}_{\text{b}} < \omega < \omega^{\mu}_{\text{max}}, \\ 
0 & {\rm otherwise}, 
\end{array} \right. 
\label{eq:spec_with_rolloff}
\end{equation}
where $A_{\mu}$ is a constant amplitude for the two error quadratures $\mu\in\{a,d\}$. 
This type of noise is frequently encountered in experimental qubit systems over a wide frequency 
range~\cite{Weissman1988,Paladino2002, Martinis2003} and naturally arises from independent bistable fluctuators~\cite{Faoro2004}.  The generality of this power spectrum in various noise processes allows us to reasonably assume the same power spectrum for both amplitude and detuning noise, despite the fact that these two noise sources have different physical origins in general and, as remarked, we take them to be independent.  Nonetheless we emphasize that our methods are independent of the specific form of the power spectrum assumed in our numerical calculations.

\begin{figure*}[htp]
\begin{center}
\includegraphics[width=\textwidth]{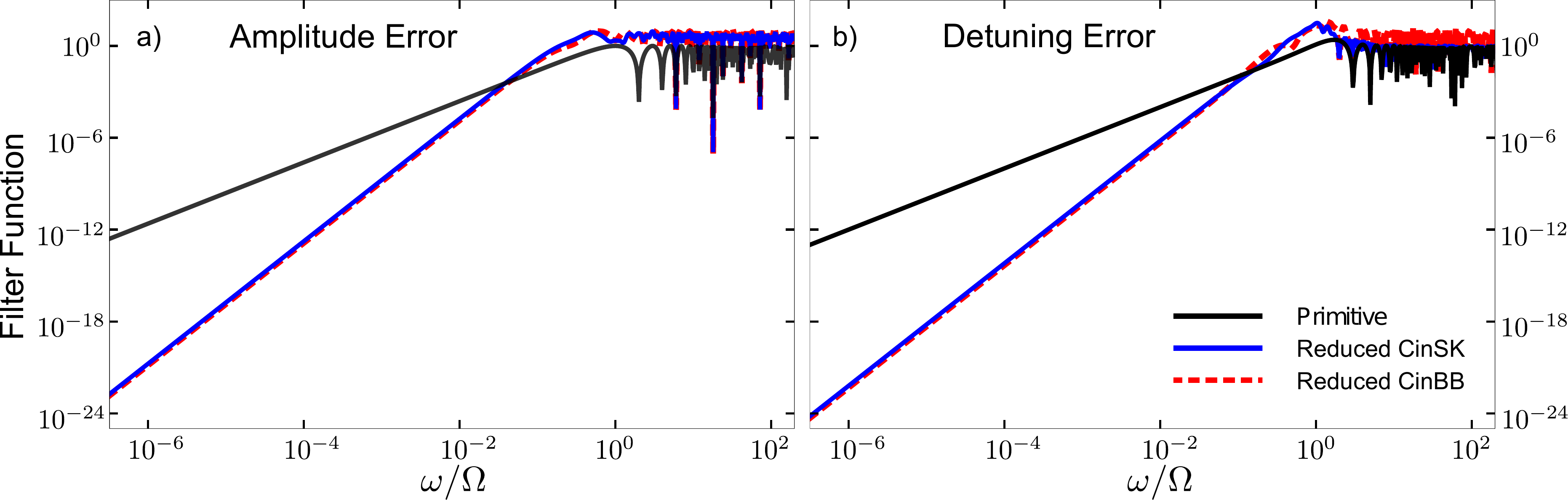}
\end{center}
\vspace*{-3mm}
\caption{(Color online) FFs as a function of dimensionless frequency for amplitude error (a) and detuning error (b) for concatenated CP sequences Reduced CinSK and Reduced CinBB. Unlike SK1, BB1, CORPSE, and DCG (see Fig. \ref{fig:pulseFF}),  these FFs scale as $\omega^4$ for both errors.}

\label{fig:simerrFF}
\end{figure*}

\begin{figure*}[htp]
\begin{center}
\includegraphics[width=\textwidth]{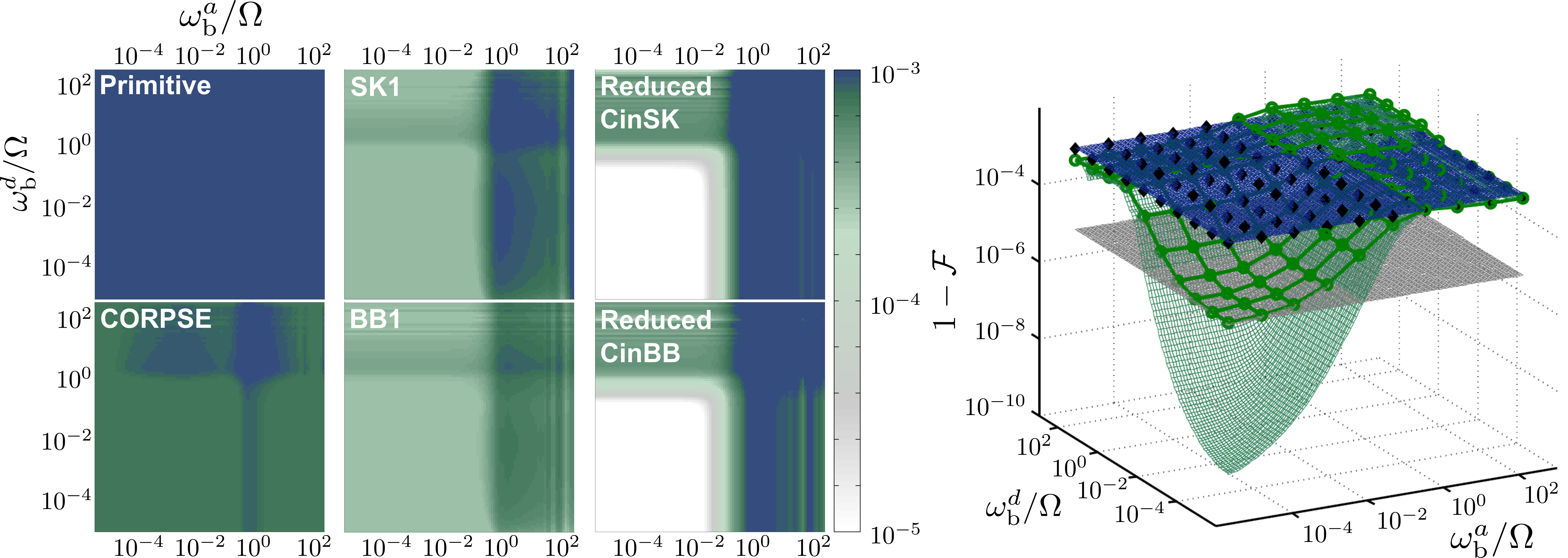}
\end{center}
\vspace*{-3mm}

\caption{(Color online) Performance of CPs under simultaneous amplitude and detuning 
noise, as a function of dimensionless frequency roll-off from $1/f$ to a $1/f^2$ spectral density, $\omega_{\text{b}}^{a}$ and 
$\omega_{\text{b}}^{d}$, respectively. 
Spectrum and control parameters as in Fig. \ref{fig:pulseFF}. Left: Analytical results for fidelity loss. For each point the FF and dc limit calculations are compared and the larger fidelity loss value is plotted. Right: Analytical (FF, green surface; dc-limit, gray surface) and numerical results (green circles and mesh) for Reduced CinBB vs. analytical (dc limit, blue surface) and numerical (black diamonds) results for a primitive pulse.}

\label{fig:simerr}
\end{figure*}

We analytically compute the fidelity loss according to~\eqn{eq:fidloss} in combination with the asymptotic error floor,~\eqn{eq:slow}.  These analytic results are compared to numerical data obtained from simulation of the Bloch vector evolution under the noisy Hamiltonian in~\eqn{eq:totham}.  Provided that the number of noise realizations, 
$N$, over which we average is large enough (typically $N \gtrsim 10^{4}$), this numeric simulation can be considered a reliable direct method for calculating the fidelity.

For the three first-order protocols studied (SK1, CORPSE, and DCG), Figs. \ref{fig:pulseFF}(c) and \ref{fig:pulseFF}(d)  show 
that as the roll-off frequency is reduced, the fidelity loss is well approximated by the combination of the FF estimate and dc limit (lines).  Vitally, both the analytic and the numerical approach 
directly reveal the robustness of CP protocols against noise fluctuations up to $\sim 0.1\Omega$. 
Detailed performance variation in the slow-noise limit stems from differences in construction of the selected gate protocols.  The DCG and CORPSE sequences both correct dc detuning noise to the first order and have first-order FF's for time-dependent errors. While for frequencies below $\sim$10\% of the Rabi frequency 
the DCG has a FF of lower magnitude than CORPSE, the specific CORPSE sequence used 
is designed to additionally minimize the residual second-order dc pre-factor~\cite{True2012} [namely, 
$c_{2}$ in Eq. (\ref{eq:slow})], which results in a dc limit of {\em X} well below the plotted fidelities. The resulting relative performance between the DCG and the CORPSE protocols further depends on the specifics of the noise power spectral density.  Similarly, the effective second-order dc error cancellation associated with BB1 means that the dc-limit does not provide a substantial contribution relative to the FF calculation for the example noise spectrum. 

Finally, we extend our analysis to include representative concatenated CP sequences 
(Table \ref{tab:comp_pulses}). We see that the FFs of the concatenated CP sequences depicted in Fig. (\ref{fig:simerrFF}) exhibit error suppression for both forms of error at low frequencies relative to a primitive pulse, in contrast to the standard CP sequences. In the presence of simultaneous noise, this leads to substantially improved performance when both noises are slow.   
 Figure \ref{fig:simerr} presents a quantitative comparison of analytical 
and numerical fidelity-loss calculations for the primitive $\pi$ pulse and for reduced CinBB, 
showing good agreement between the two approaches.  For this two-parameter 
compensating sequence, the constant-error DC fidelity limit may be seen to arise due to a cross-term of 
the two noise sources,  namely, 
$$1-\mathcal{F} =c_{1,1} \< \beta_a(0)^{2}\beta_d(0)^{2} \>, $$ 
\noindent 
where $c_{m+1,n+1}$ is the cross-term equivalent of $c_{m+1}$ for single noise sources in Eq. (\ref{eq:slow}).  
As the data show, the resulting dc limit matches the fidelity loss in the very-low-frequency regime 
for the reduced CinBB sequence.

Our numerical calculations validate the insights provided by the analytic FF formalism 
and demonstrate that, in combination with the calculated dc error floor, the first-order FF is an 
effective tool for predicting single-qubit control performance 
in the presence of time-dependent noise.  The analytic approach comes with an additional benefit, 
however, in terms of computational efficiency; the numerical calculations of fidelity loss under 
time-dependent noise are in fact significantly more computationally intensive than the FF 
approach \cite{Remark}.  
While this is beyond our current purpose, this advantage is likely to become even more 
dramatic in more complex control scenarios, in particular, including multiple qubits. 

\section{Conclusion}

We have shown that CP sequences originally designed to compensate only for static
control errors may be successfully employed for non-Markovian time-dependent control 
and/or environmental errors as well.  Our numeric and analytic results demonstrate 
that these sequences are robust against noise fluctuations up to $\sim$10\% of the 
control frequency, a surprisingly high value.  
In addition to substantially expanding the practical significance of open-loop quantum control protocols, 
our analysis further establishes the utility of FFs as a unifying and computationally efficient framework 
for estimating and understanding the performance of coherent control protocols under realistic noise spectra. Furthermore, we have shown that at least for the single-qubit setting under consideration, slow noise can be 
accurately modeled by a dc-limit approximation that can be combined with the FF approach to accurately 
estimate control performance over a broader frequency range.  

Altogether, our results show that, in combination, CP and DCG protocols provide 
experimentalists with a viable toolkit capable of meeting a variety of constraints, including the presence of 
colored time-dependent control noise. We further expect that the geometric picture we have developed, in conjunction with the FF approach, may prove instrumental for finding new CPs which are resilient to 
specific noise spectra.

\section*{ACKNOWLEDGMENTS} 

It is a pleasure to thank Harrison Ball, J. True Merrill, and Alexander Soare for helpful comments. 
This work was supported by the Office of the Director of National Intelligence IARPA
through ARO Contract No. W911NF-10-1-0231 and Department of Interior Contract No. D11PC20167, ARC Centre for Engineered Quantum Systems Grant No. CE110001013, the U.S. ARO under Contract No. W911NF-11-1-0068, and Lockheed Martin Corporation.

The views and conclusions contained herein are those of the authors and should not be interpreted as necessarily representing the official policies or endorsements, either expressed or implied, of IARPA, DoI/NBC, or the U.S. Government.

\appendix
\section*{APPENDIX: TRAPEZOIDAL PULSES}
\label{app:shape}
In actual experiments, the pulse shape deviates from the ideal square-pulses under which CPs are derived. This is often done on purpose when, for example, Gaussian pulses or Blackman pulses are used to limit the spectral bandwidth of the control \cite{Thom2013}. This also occurs accidentally due to bandwidth limitations of the instrument resulting in fast amplitude fluctuations or slow turn-off times. Although the FF formalism as described in Sec. \ref{subs:analysis} assumes piecewise-constant control, continuous pulse-modulation profiles can be analyzed by a discrete time-step approximation. We apply this approximation to examine the effect of pulse shape on CP FFs for amplitude and detuning noise.

 We expect that the FF of amplitude noise CPs will be weakly dependent on the pulse shape since amplitude noise, unlike detuning noise, commutes with the control pulse.  In fact, using the error model of Eq. (\ref{eq:totham}), the FF is pulse-shape independent if the total pulse time is the same as the square pulse it replaces. CPs for amplitude noise were developed assuming that the error is proportional to the control (multipicative noise). This noise can be modeled in our formalism by replacing $\beta_a(t)$ in Eq. (\ref{eq:totham}) with $\Omega_l/\Omega_{max}\beta_a(t)$. We note that additive and multiplicative error models are equivalent for the constant $\Omega$ pulses considered in the main text. In the case of multiplicative noise, static error correction only requires the rotation angle be constant. On the other hand, detuning noise does not commute with the control, and as a result the pulse shape can have a significant effect.   

\begin{figure*}[htp]
\begin{center}
\includegraphics[width=\textwidth]{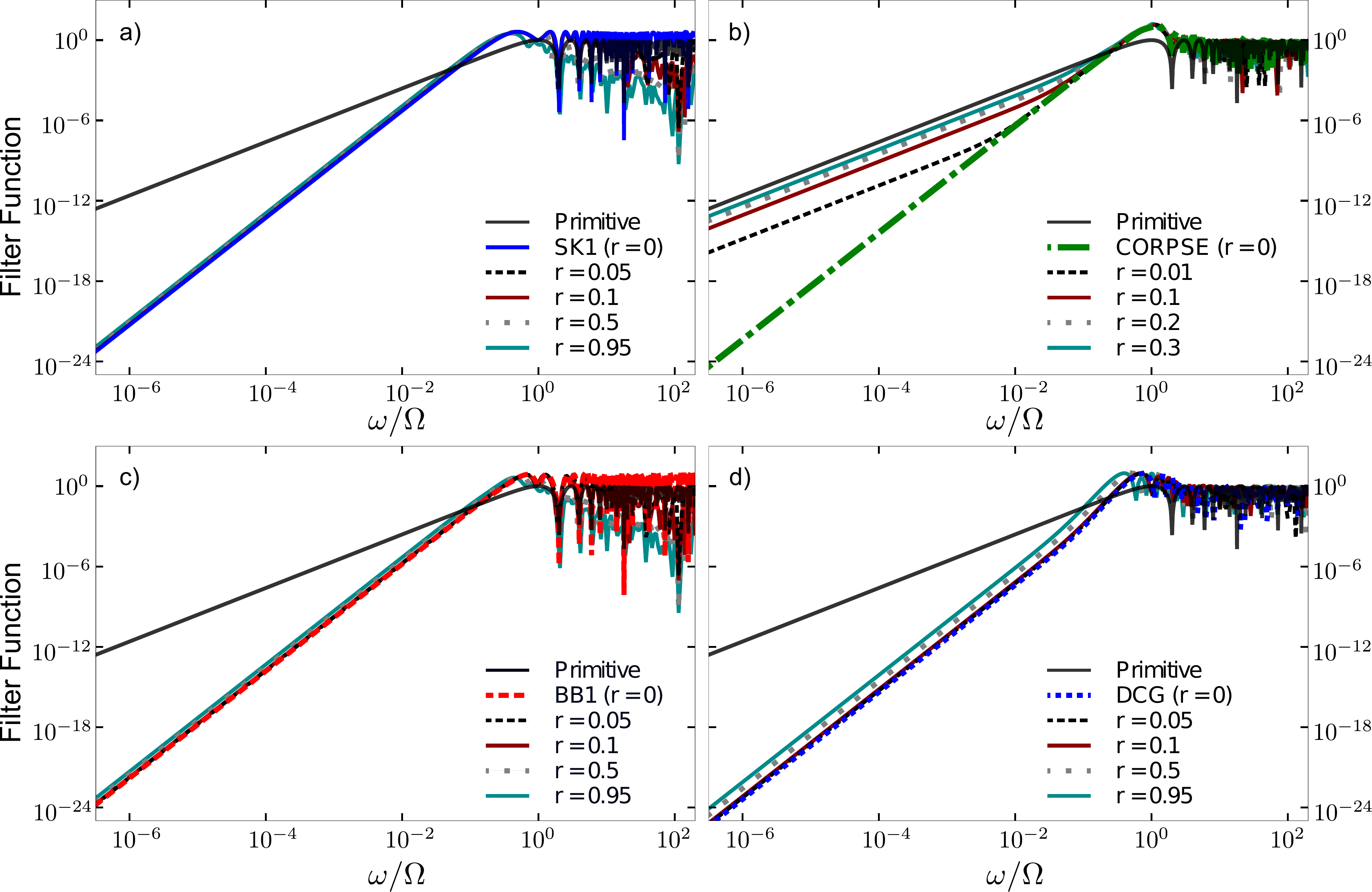}
\end{center}
\vspace*{-3mm}
\caption{(Color online) FFs as a function of dimensionless frequency for SK1 [Panel a)] and BB1 [Panel c)] in the presence of multiplicative amplitude noise and for CORPSE [Panel b)] and DCG [Panel d)] in the presence of detuning noise. The CPs are constructed from trapezoidal pulses with ramp time $r$ in units of $\pi/\Omega$.}
\label{fig:trapeza}
\end{figure*}

As an example, we examine trapezoidal pulses where the $k$-th pulse is ramped up to $\Omega_{k}$ in a time $r$, held for a time $w$, and then ramped down in a time $r$. The total pulse time is $w+2r$ and $w+r$ is held constant to preserve the rotation angle. For the CPs studied here, $\Omega_{k}=\Omega$. BB1 and SK1 are designed assuming a systematic and proportional error in the rotation angle. This is preserved for multiplicative amplitude noise, and we see that the FF form is maintained (Fig. \ref{fig:trapeza}). There is an increase in the magnitude of the FF in the small-$\omega$ region due to the increase in the overall sequence length in time. 

CORPSE is designed under the assumption of square pulses and the detuning is additive. Consequently, trapezoidal pulses do not perfectly remove the first-order error using the rotation angles of CORPSE.  This changes the asymptotic behavior of the FF and we see a bend corresponding to the residual $\omega^2$ term due to imperfect error cancellation [Fig. \ref{fig:trapeza}b]. The bend occurs at lower frequencies as the control approaches a square profile. 

In contrast, the design of $\pi$ DCG does not assume square pulses \cite{DCG1}. The static error cancellation will occur if the first and the third pulses have the same time-dependent control profile applied for a total time $T$ and the second pulse has the {\em stretched and scaled} control profile applied for time $2T$. The parameters for the first and second trapezoidal  pules are related as follows: $2r_1 = r_2 $, $2 w_1=w_2$, and $\Omega_{1}/2=\Omega_{2}$  The FF form at small $\omega$ remains unchanged and the magnitude again increases with overall sequence length [Fig. \ref{fig:trapeza}d].

In practice, if square pulses are not an adequate approximation, then CORPSE should not be used.  Instead a DCG should be chosen or one can derive a CORPSE-like sequence using soft pulses to achieve similar slow-noise cancellation \cite{Sengupta2005}.

\bibliographystyle{apsrev}

\begin{thebibliography}{31}
\bibitem{Remark} For our current analysis, numerical calculation of the fidelity numerically requires the generation of $N \gtrsim 10^{4}$ noise trajectories for each spectral density. The FF and dc-limit calculations of fidelity require only two numerical integrals per spectral density.
\end{thebibliography}

\end{document}